\documentstyle[sprocl]{article}

\input{psfig}
\def\NIM{\em Nucl. Instrum. Methods}

\def\NPB{{\em Nucl. Phys.} B}
\def\PLB{{\em Phys. Lett.}  B}
\def\PRL{\em Phys. Rev. Lett.}
\def\PRD{{\em Phys. Rev.} D}

\def\be{\begin{equation}}
\def\nn{\noindent}

\def\eg{{\it e.g.}}
\def\etc{{\it etc}}
\def\etal{{\it et al.}}
\def\ee{\end{equation}}
\def\bea{\begin{eqnarray}}
\def\eea{\end{eqnarray}}

\begin{document}

\rightline{\vbox{\halign{&#\hfil\cr
SLAC-PUB-8204\cr
July 1999\cr}}}
\vspace{0.8in}

\title{{TESTS OF LOW SCALE QUANTUM GRAVITY IN $e^-e^-$ AND $\gamma \gamma$ 
COLLISIONS}
\footnote{To appear in the {\it Proceedings of the World-Wide Study of 
Physics and Detectors for Future Linear Colliders(LCWS99)}, Sitges, 
Barcelona, Spain, 28 April-5 May 1999}
}

\author{ {T.G. RIZZO}
\footnote{Work supported by the Department of Energy, 
Contract DE-AC03-76SF00515}
}

\address{Stanford Linear Accelerator Center,\\
Stanford University, Stanford, CA 94309, USA}


\maketitle\abstracts{Arkani-Hamed, Dimopoulos and Dvali have recently 
proposed that gravity may become strong at energies near 1 TeV due to the 
existence of large extra dimensions thus `removing' the hierarchy problem. 
In this talk we examine the exchange of towers of Kaluza-Klein gravitons 
and their influence on Moller scattering as well as the production of pairs 
of massive gauge bosons in $\gamma \gamma$ collisions. These tower exchanges 
lead to a set of new dimension-8 operators that can significant alter 
the Standard Model expectations for these processes. In the case of 
$\gamma \gamma$ collisions, the role of polarization for both the initial 
state photons and the final state gauge bosons in improving sensitivity to 
graviton exchange is emphasized.}

Recently, ~Arkani-Hamed, ~Dimopoulos and ~Dvali(ADD)~{\cite {nima}} have 
proposed an interesting solution to the hierarchy problem. ADD 
hypothesize the existence of $n$ additional large spatial dimensions in 
which gravity (and perhaps Standard Model singlet fields) can live, called 
`the bulk', whereas all of the fields of the Standard Model(SM) are 
constrained to lie on `the wall', which is our 
conventional 4-dimensional world. In such a theory the hierarchy is 
removed by postulating that the string or effective Planck scale in 
the bulk, $M_s$, is not far above the weak scale, \eg, a few TeV. Gauss' Law 
then provides a link between the values of $M_s$, the conventional 
Planck scale $M_{pl}$, and the size of the compactified extra dimensions, $R$: 
$M_{pl}^2 \sim R^nM_s^{n+2}$
where the constant of proportionality depends upon the geometry of the 
compactified dimensions. If $M_s$ 
is near a TeV then $R\sim 10^{30/n-19}$ meters; for separations between two 
masses less than $R$ the gravitational force law becomes $\sim 1/r^{2+n}$. 
For $n=1$, $R\sim 10^{11}$ meters and is thus 
excluded~{\cite {ran}}, but, 
for $n=2$ one obtains $R \sim 1$~mm, which is at the edge of the sensitivity 
for existing experiments~{\cite {test}}. 
Astrophysical arguments based on supernova cooling and cosmological 
arguments seem to require~{\cite {astro} that 
$M_s>100$ TeV for $n=2$, but allow $M_s\sim 1$ TeV for $n>2$.

The detailed phenomenology of the ADD model has begun to be explored for a 
wide ranging set of processes in a growing series of recent 
papers~{\cite {pheno}} where it has been shown that the 
ADD scenario has two basic classes of collider tests. In the first class, 
a K-K tower of gravitons can be emitted during a decay or scattering process 
leading to a final state with missing energy. The rate for such processes is 
strongly dependent on the number of extra dimensions as well as the exact 
value of $M_s$. 
In the second class, which we consider here~{\cite {tgr}}, the exchange of a 
K-K graviton tower between SM fields can lead to almost 
$n$-independent modifications to conventional cross sections and 
distributions or they can possibly lead to new interactions. The exchange of 
the graviton K-K tower leads to a set of effective color and flavor singlet 
contact interaction operator of dimension-eight with the overall scale set by 
the cut-off in the tower summation, $\Lambda$, which naively  
should be of comparable magnitude to $M_s$. Thus one introduces a 
universal overall order one coefficient for these 
operators, $\lambda$ ( whose value is unknown but can be approximated by a 
constant which has conventionally been 
set to $\pm 1$) with $\Lambda$ being replaced by $M_s$.

Linear colliders will provide the opportunity to make precision measurements 
of a number of elementary processes making small deviations from SM 
expectations due to new physics easily observable. It is well known that 
Moller scattering is particularly sensitive to both contact interactions as 
well as new neutral gauge bosons so we may expect reasonable sensitivity to 
graviton exchange. 
The shifts in the angular distribution due to gravity are shown in a sample 
case in Fig. 1; additional information can be obtained from the left-right 
polarization asymmetry, $A_{LR}$. Following now-standard 
procedures~{\cite {tgr}} we 
can estimate the search reach for $M_s$ for a fixed sign of $\lambda$ 
assuming a given integrated luminosity 
and beam polarization via a Monte Carlo approach. 
In particular the search reach is obtained by fitting to the total 
number of events, the shape of the 
angular distribution and the angle-dependent values of $A_{LR}$. 
The result of this approach is also shown in Fig.1 
together with the corresponding 
reaches obtained from Bhabha scattering and from combining 
multiple final states using the process $e^+e^- \to f\bar f$ [where $f=\mu, 
\tau, ~b, ~c, ~t$ \etc.] as obtained by 
Hewett~{\cite {pheno}}. While for typical luminosities the inclusive analysis 
leads to an $M_s$ reach of $\sim (6-7)\sqrt s$, the corresponding reach for 
Moller scattering is $\sim (5-6)\sqrt s$.

\vspace*{-0.5cm}
\nn
\begin{figure}[htbp]
\centerline{
\psfig{figure=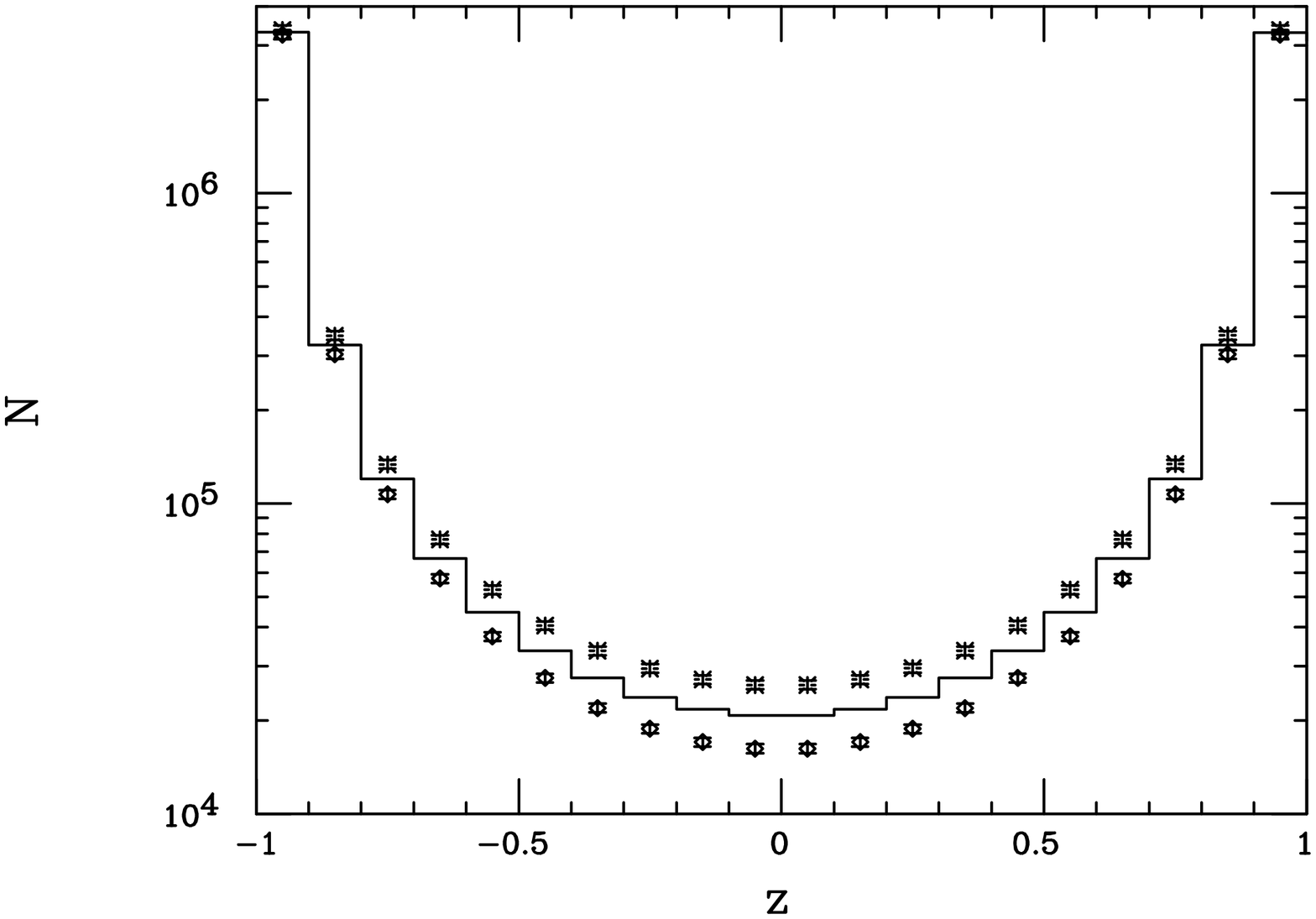,height=5.8cm,width=6.8cm,angle=-90}
\hspace*{-5mm}
\psfig{figure=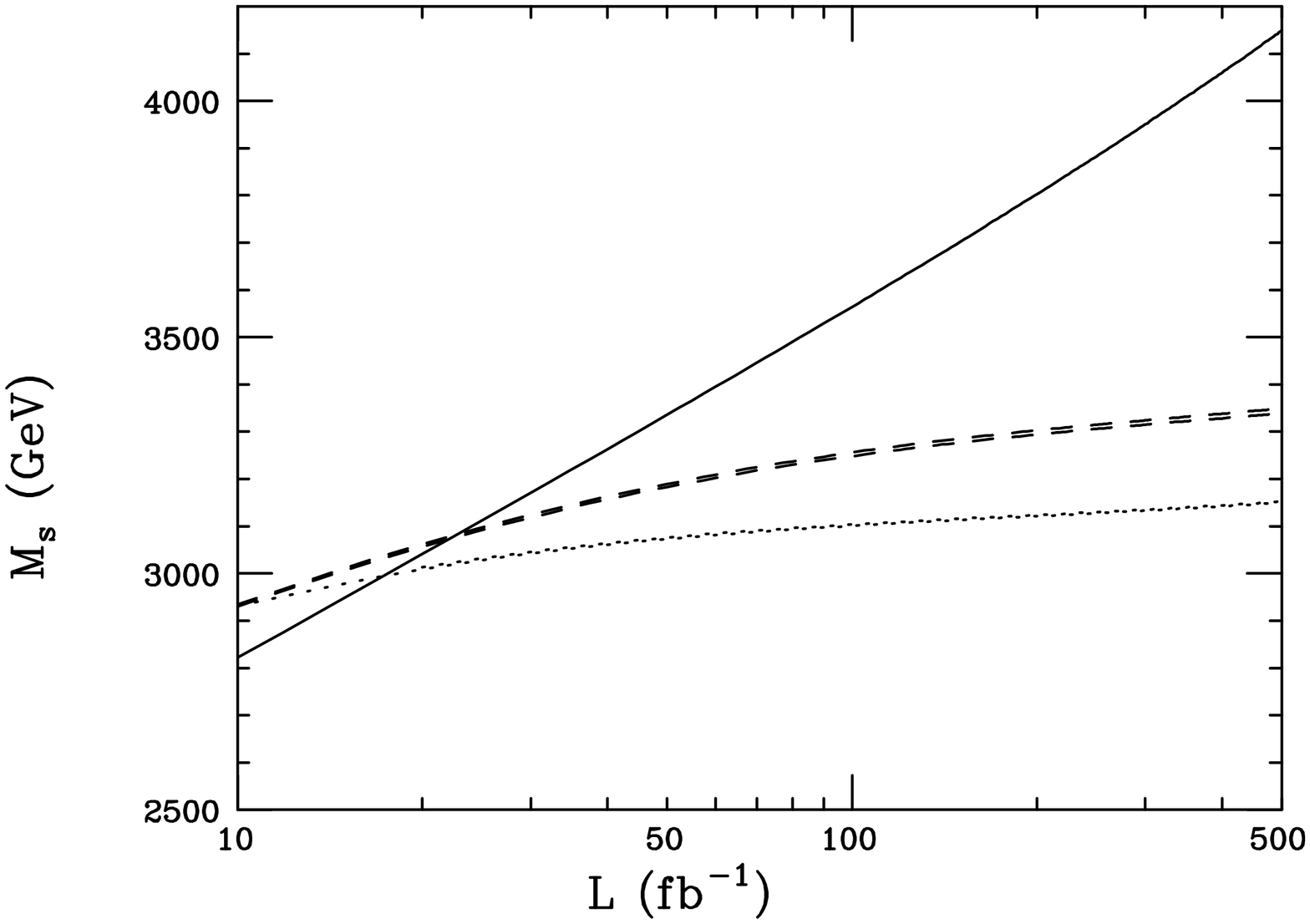,height=5.8cm,width=6.8cm,angle=-90}}
\vspace*{-1cm}
\caption{(Left) Deviation from the expectations of the SM(histogram) for Moller
scattering at a 500 GeV $e^+e^-$ collider for the number of events per 
angular bin, $N$, as a function of 
$z=\cos \theta$ assuming $M_s$=1.5 TeV. The two sets of data points 
correspond to the choices $\lambda=\pm$ 1 and an assumed integrated 
luminosity of $75~fb^{-1}$.
(Right) Search reaches for $M_s$ at a 500 GeV $e^+e^-/e^-e^-$ collider as 
a function of the integrated luminosity for Bhabha(dashed) and Moller(dotted) 
scattering for either sign of the parameter $\lambda$ in comparison to the 
`usual' search employing $e^+e^-\to f\bar f$, inclusively.}
\end{figure}
\vspace*{0.4mm}

$\gamma \gamma$ collisions may be possible at future 
$e^+e^-$ linear colliders by the use of Compton backscattering of low energy 
laser beams~{\cite {telnov}}. The backscattered laser photon spectrum, 
$f_\gamma(x={E_\gamma \over E_e})$, is far from being monoenergetic and is 
cut off above $x_{max}\simeq 0.83$ implying that the photons are significantly 
softer than their parent lepton beam energy. The shape of the spectrum as well 
as the energy dependence of the polarization of the resulting hard photon 
depends upon the polarizations of both the laser and initial electron beam. 
In what follows we will label the six independent polarization 
possibilities by the corresponding signs of the electron and laser 
polarizations as $(P_{e1},P_{l1},P_{e2},P_{l2})$, For example, the 
configuration $(-++-)$ corresponds to $P_{e1}=-0.9$, $P_{l1}=+1$, $P_{e2}=0.9$ 
and $P_{l2}=-1$. Clearly some of these polarization 
combinations will be more sensitive to the effects of K-K towers of gravitons 
than others.

\vspace*{-0.5cm}
\nn
\begin{figure}[htbp]
\centerline{
\psfig{figure=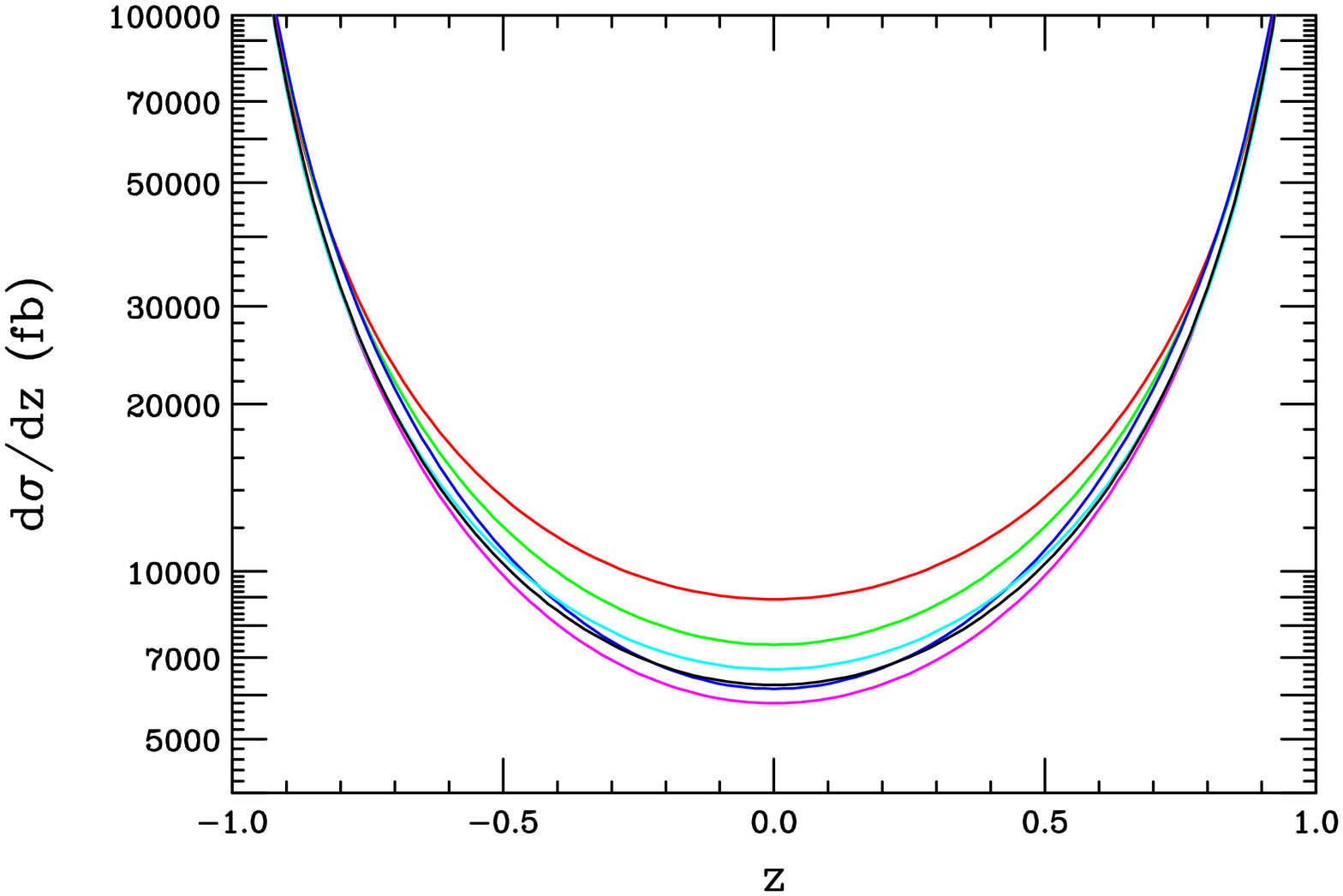,height=4.7cm,width=5.7cm,angle=90}
\hspace*{5mm}
\psfig{figure=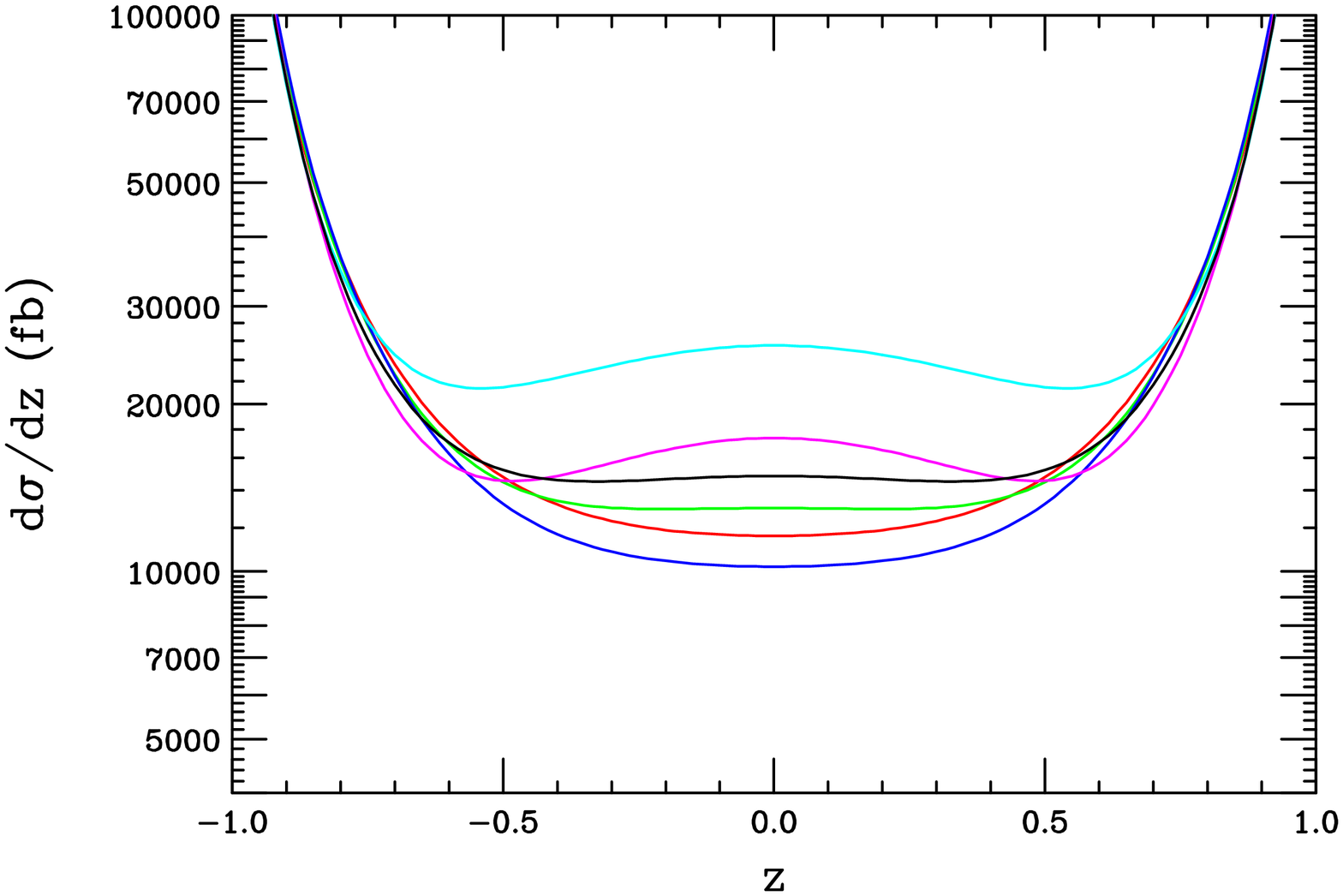,height=4.7cm,width=5.7cm,angle=90}}
\vspace*{-0.2cm}
\caption{Differential cross section for $\gamma\gamma \to W^+W^-$ at a 1 TeV 
$e^+e^-$ collider for (left)the SM and with $M_s=2.5$ TeV 
with (right)$\lambda=1$. In the left panel, from top to bottom in the center 
of the figure the helicities are $(++++)$, $(+++-)$, $(-++-)$, $(++--)$, 
$(+---)$, and $(+-+-)$; in the right panel they are  
$(-++-)$, $(+-+-)$, $(+++-)$, $(+---)$, $(++++)$, and $(++--)$.}
\end{figure}
\vspace*{0.4mm}

$\gamma \gamma$ collisions offer a unique and distinct window on the 
possibility of new physics in a particularly clean environment. 
Unlike particle production in 
$e^+e^-$ collisions, however, $P$, $C$, plus the Bose symmetry of the 
initial state photons forbids the existence of non-zero values, at the 
tree level, for either forward-backward angular asymmetries or left-right 
forward-backward polarization asymmetries. These were both found to be 
powerful tools in probing for K-K graviton tower exchanges in 
the $e^+e^-$ initiated channels~{\cite {pheno}}.
By analogy with the $e^+e^-\to f\bar f$ analysis above one can examine the 
corresponding $\gamma \gamma \to f\bar f$ process. In the case the best reach 
is obtained by combining the three lepton flavors, $t\bar t$ and $jj$ final 
states. This has been done in the 
case of unpolarized beams~{\cite {tgr}} and yields a search reach of 
$(4-5)\sqrt s$. 
For any given choice of the initial state laser and electron polarizations, 
labelled by $(a,b)$ below, we can immediately write down the appropriate 
cross section by folding in the corresponding photon fluxes and integrating:
\begin{equation}
{d\sigma^{ab}\over {dz}}=\int ~dx_1~\int~dx_2~f^a_\gamma(x_1,\xi_1)
f^b_\gamma(x_2,\xi_2)\Bigg[{{1+\xi_1\xi_2}\over {2}}
{d\hat \sigma_{++} \over {dz}}+{{1-\xi_1\xi_2}\over {2}}
{d\hat \sigma_{+-} \over {dz}}\Bigg]\,.
\end{equation}
The $++$ and $+-$ labels on the 
subprocess cross sections indicate the appropriate values of $\lambda_{1,2}$ 
to chose in their evaluation.

\vspace*{-0.5cm}
\nn
\begin{figure}[htbp]
\centerline{
\psfig{figure=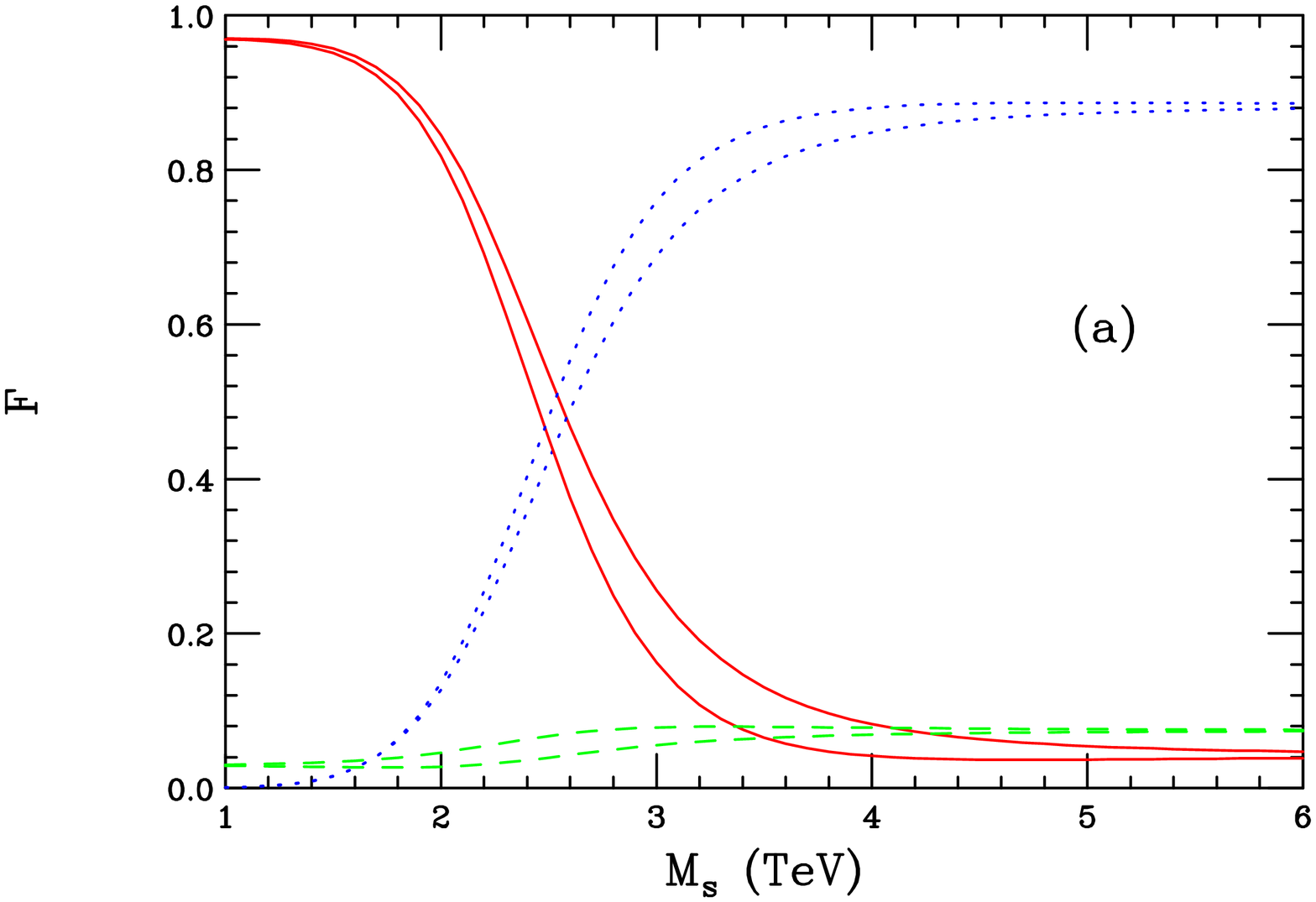,height=4.7cm,width=5.7cm,angle=90}
\hspace*{5mm}
\psfig{figure=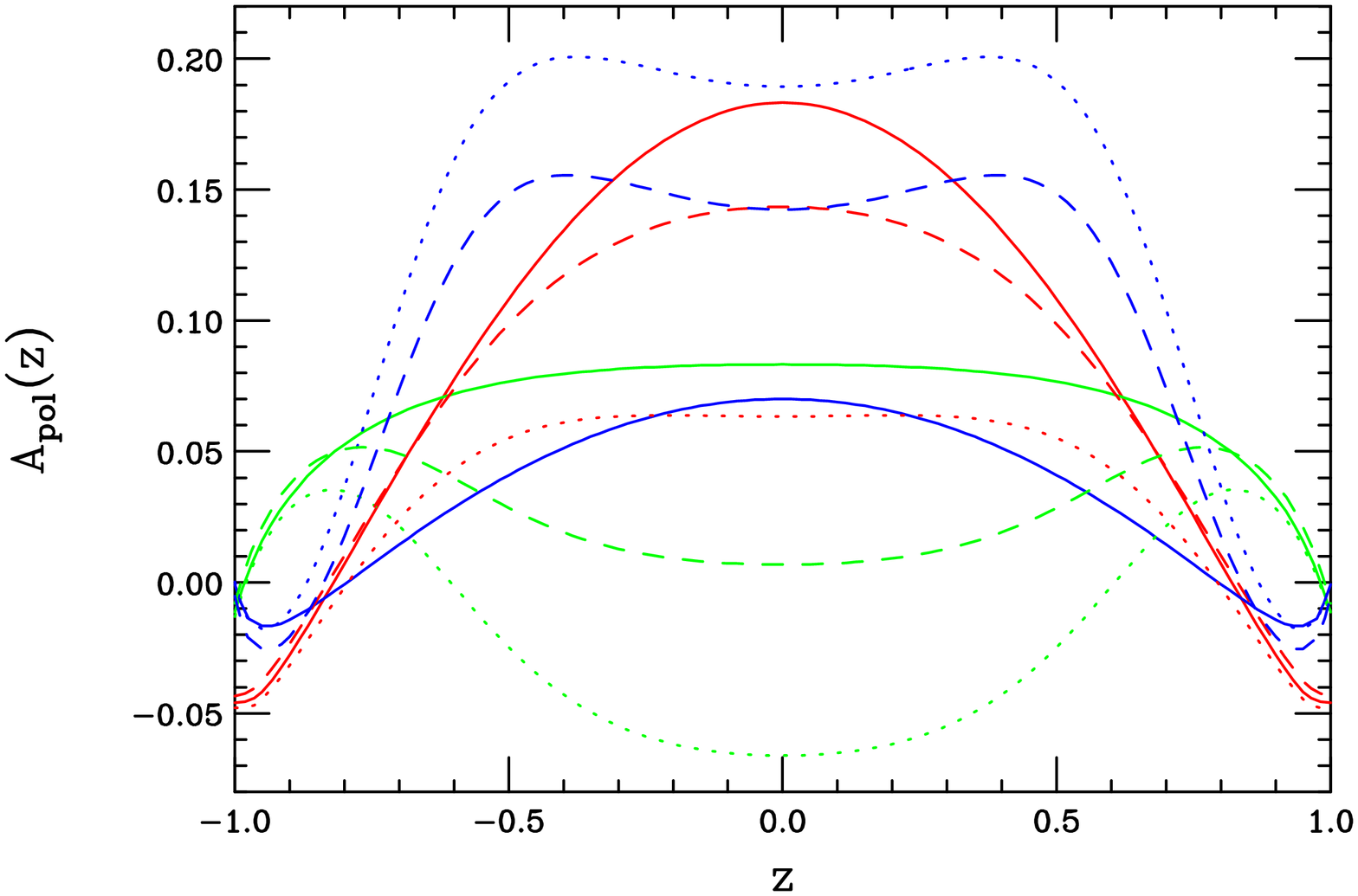,height=4.7cm,width=5.7cm,angle=90}}
\vspace*{-0.2cm}
\caption{(Left) Fraction of LL(solid), TL+LT(dashed) and TT(dotted) $W^+W^-$ 
final states after angular cuts for the process $\gamma \gamma \to W^+W^-$ at a 
1 TeV $e^+e^-$ collider as a function of $M_s$ for either sign of $\lambda$. 
The initial state polarization is $(-++-)$. (Right) Differential polarization 
asymmetries for $\gamma \gamma \to W^+W^-$ at a 1 TeV 
$e^+e^-$ collider for the SM(solid) as well with graviton tower exchange with 
$M_s$=2.5 TeV with $\lambda=\pm 1$(the dotted and dashed curves). We label the 
three cases shown by the first entry in the numerator in the definition of 
$A_{pol}$. Red represents an initial polarization of $(++++)$, green is for the 
choice $(+++-)$ and blue is for the case $(-++-)$.}
\end{figure}
\vspace*{0.4mm}

Perhaps the most interesting channels are $\gamma \gamma \to VV$ where $V$ is 
either $Z$ or $W$. While the first process only occurs at loop level in the 
SM (though it can now be mediated by tree level graviton exchange), the 
second has a very large tree level cross section. In either 
case many observables which may be sensitive to graviton tower exchange 
can be constructed in addition to angular distributions 
due to ($i$) the existence of the six distinct initial $\gamma \gamma$ 
polarization states allowing us to construct three different polarization 
asymmetries and ($ii$) the 
fact that the polarizations of the final state vector bosons can be measured 
through angular correlations of their decay products. The shifts in the $WW$ 
angular distributions are shown in Fig. 2. In the SM, the
$\gamma \gamma \to W^+W^-$ reaction takes place through $t$- and $u$-channel 
$W$ exchanges as well as a $\gamma \gamma W^+W^-$ four-point interaction. The 
$t$- and $u$-channel exchanges thus lead to a sharply rising cross section 
in both the forward and backward directions.  Note that in the SM there is no 
dramatic sensitivity to the initial state 
polarizations and all of the curves have roughly the same shape. 
In all cases the total cross section, even after generous angular cuts, is 
quite enormous, of order $~\sim 100$ pb, providing huge statistics to look 
for new physics influences.  
When the graviton terms are turned on there are several effects. First, all 
of differential cross section distributions become somewhat more shallow, 
but there is little change in the forward and backward 
directions due to the dominance of the SM poles. Second, there is now a 
clear and distinct sensitivity to the initial state 
polarization selections. In some cases, particularly for the $(-++-)$ and 
$(+-+-)$ helicity choices, the differential cross section increases 
significantly for angles near $90^o$ taking on an m-like shape. This shape is, 
in fact, symptomatic of the spin-2 nature of the K-K graviton tower exchange 
since a spin-0 exchange leads only to a flattened distribution. 
In addition to a significant modification to the angular distribution, the K-K
graviton tower exchange 
influences on the polarizations of the two $W$'s in the final state. In the SM, 
independent of the initial electron and laser polarizations, the final state 
$W$'s are dominantly transversely polarized. Due to the nature of the spin-2 
graviton exchange, the K-K tower leads to a final state where both $W$'s are 
completely longitudinally polarized. To see this, we show in Fig. 3
the polarization fractions of the two $W$'s as a function of $M_s$ at a 1 TeV 
collider. Here we see that the fraction of final states where both $W$'s are 
longitudinal, denoted by $LL$, starts out near unity but falls significantly 
in the $M_s=2.5-3$ TeV region giving essentially the SM results above 
$M_s \simeq $5.5-6 TeV. The reverse situation is observed for the case where 
both $W$'s are transversely polarized, denoted by $TT$. 

The three polarization asymmetries are also shown in Fig. 3. Note that as 
$z\to \pm 1$ the SM dominates due to the large magnitude of the $u$- and 
$t$-channel poles. Away from the poles the three asymmetries all show a 
significant sensitivity to the K-K tower of graviton exchange. Although these 
asymmetries are not very big the large statistics of the data samples 
obtainable for this channel indicate that they will be very well determined
since many systematic errors will also cancel in forming the cross section 
ratios. It is clear from the discussion above that there are a large number of 
observables that can be combined into a global fit to probe very high values 
of $M_s$ in comparison to the collider energy.
It should be noted however that due to the large statistics available 
the eventually determined discovery reach for $M_s$ using the 
$\gamma \gamma \to W^+W^-$ process will strongly depend on the size and 
variety of the experimental systematic errors. 
The results of performing this fit for $\lambda=1$ and for the six possible 
initial state polarizations are displayed in Fig. 4
which displays the reach as a function of the integrated luminosity. (The 
results for $\lambda=-1$ are almost identical.) 
Note both the strong sensitivity of the reach to the initial electron and 
laser polarizations as well as the 
large values obtainable particularly for the $(-++-)$ choice. In this 
particular case with 100 $fb^{-1}$ of integrated luminosity the discovery 
reach is almost $11\sqrt s$ for either sign of $\lambda$, which is greater 
than any other K-K graviton 
exchange process so far examined. 

\vspace*{-0.5cm}
\nn
\begin{figure}[htbp]
\centerline{
\psfig{figure=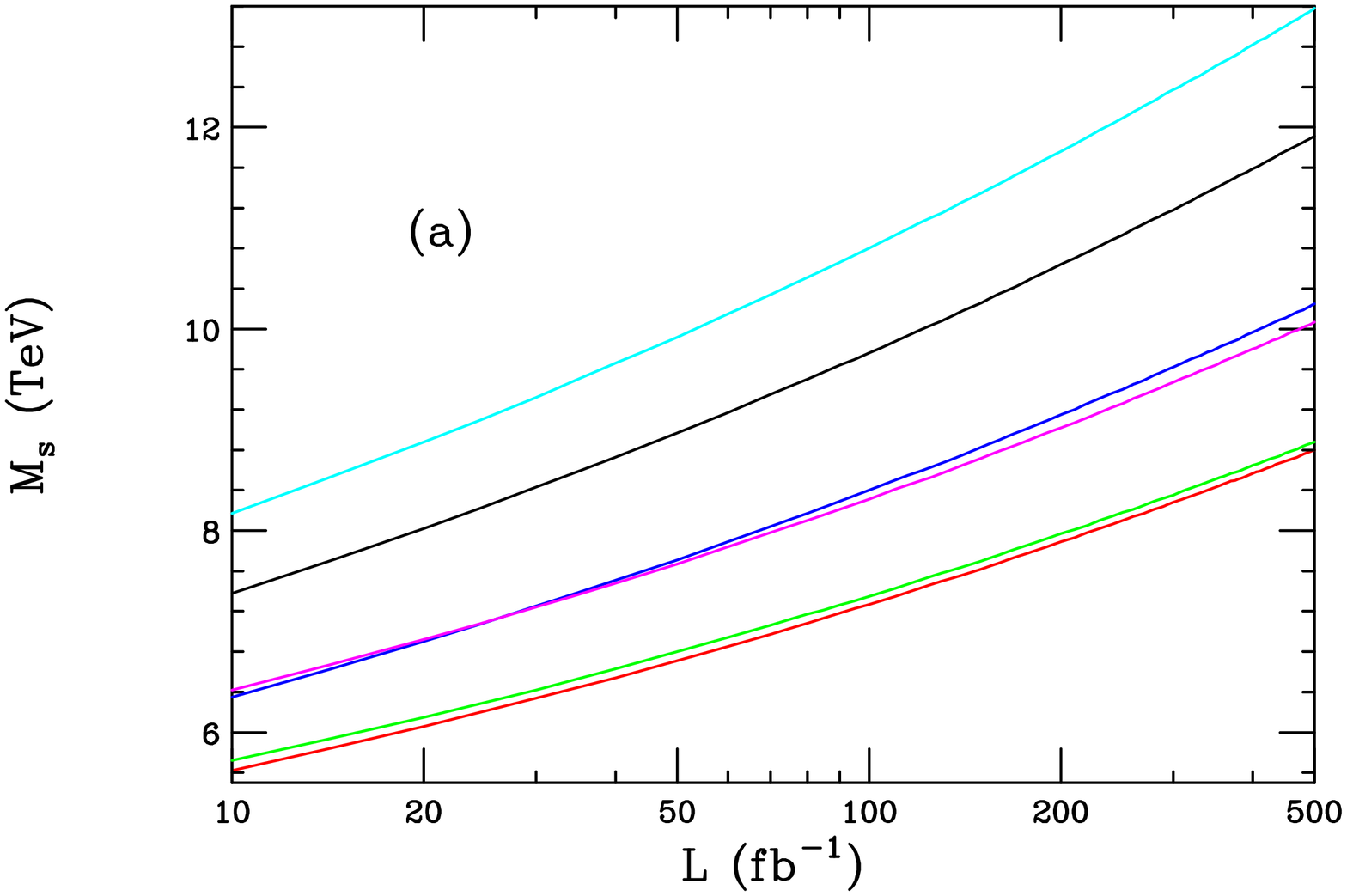,height=4.7cm,width=5.7cm,angle=90}
\hspace*{5mm}
\psfig{figure=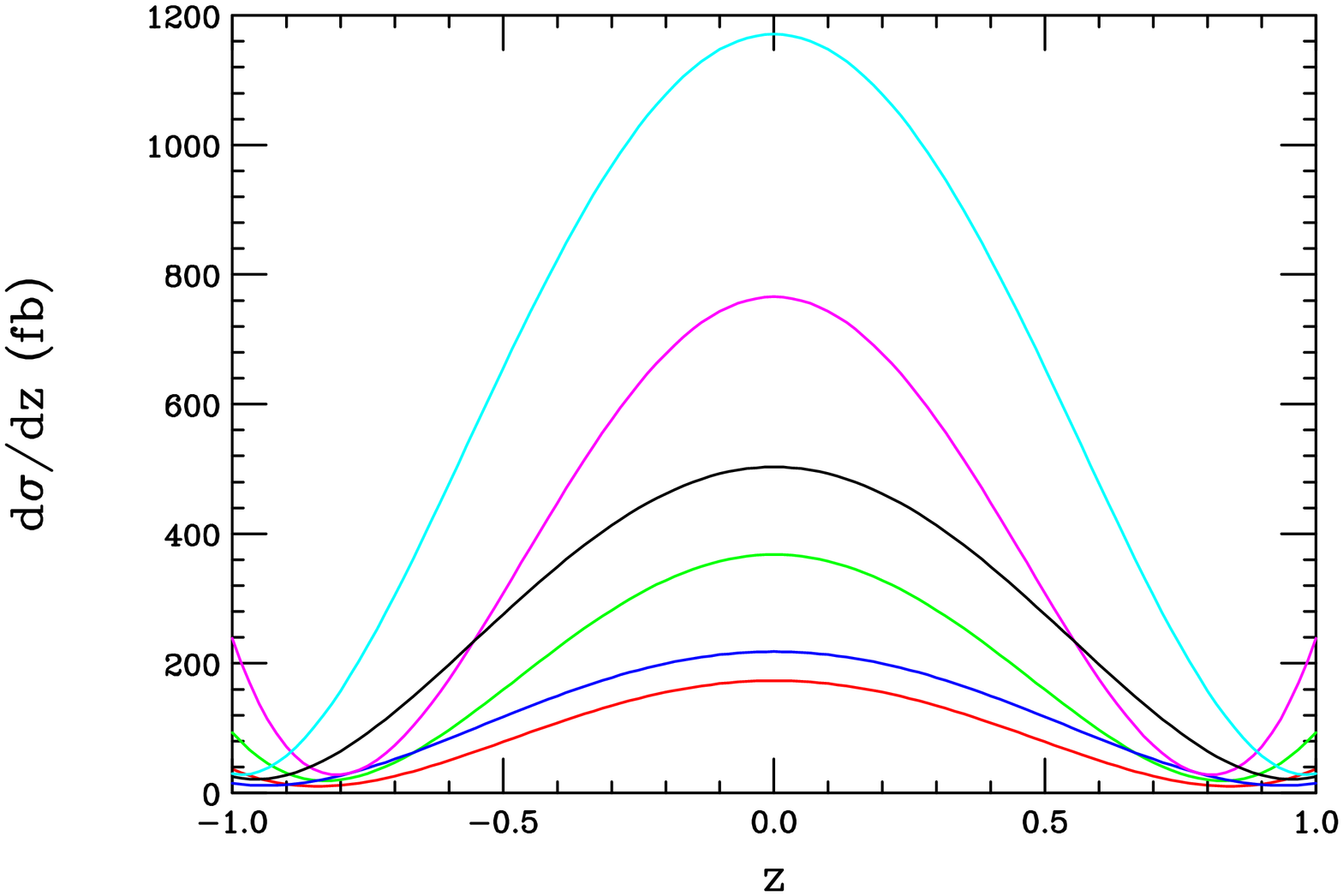,height=4.7cm,width=5.7cm,angle=90}}
\vspace*{-0.2cm}
\caption{(Left) $M_s$ discovery reach for the process 
$\gamma \gamma \to W^+W^-$ at a 1 TeV $e^+e^-$ collider as a function of the 
integrated luminosity for the different initial state polarizations assuming 
$\lambda=1$. From top to bottom on the right hand side of the figure the 
polarizations are $(-++-)$, $(+---)$, $(++--)$, $(+-+-)$, $(+---)$,  
and $(++++)$. (Right) Differential cross section for $\gamma \gamma \to ZZ$ 
at a 1 TeV 
$e^+e^-$ collider due to the exchange of a K-K tower of gravitons assuming 
$M_s=3$ TeV. From top to bottom in the center of the figure the initial state 
helicities are $(-++-)$, $(+-+-)$, $(+---)$, $(+++-)$, $(++--)$, $(++++)$.}
\end{figure}
\vspace*{0.4mm}

The process $\gamma \gamma \to ZZ$ does not occur at the tree level in the SM 
or MSSM. At the one loop level in the SM the dominant 
contribution arises from $W$ and 
fermion box diagrams and triangle graphs with $s$-channel Higgs boson 
exchange. This would seem to imply 
that this channel is particularly suitable for looking for new physics effects 
since the SM and MSSM rates will be so small due to the loop suppression. 
The SM cross section (which peaks in the forward and backward directions), 
after a cut of $|z|<0.8$, is found to be $\sim 80$ fb and almost purely 
transverse away from Higgs boson resonance peaks. 
In the case of the ADD scenario the tree level K-K graviton tower contribution 
is now also present. Neglecting the loop-order SM 
contributions for the moment we obtain the polarization-dependent 
differential cross sections shown in Fig. 4. Note that since this is 
the pure K-K graviton tower term there is no dependence here on the sign of 
$\lambda$. This cross section is found to scale with $s$ and $M_s$ as 
$\sim s^3/M_s^8$ and in contrast to the SM case is observed to peak at $90^o$. 
A short analysis shows that essentially all of the $Z$'s in the final 
state are completely longitudinal with the $LL$ fraction being 
$\sim 99\%$ for the six possible initial state polarizations. 
Although a detailed study of the loop-induced SM-graviton tower exchange 
interference terms have not yet been performed it is difficult to see how the 
search reach in this channel can exceed $\simeq 5$ TeV given the small 
magnitudes of the cross sections involved. Thus this process is not 
competitive with $\gamma \gamma \to WW$. 

Signals for an exchange of a Kaluza-Klein tower of gravitons in the ADD 
scenario of low scale quantum gravity appear in many complementary channels 
{\it simultaneously} at various colliders. Such signatures for new physics are 
rather unique and will not be easily missed.

\section*{Acknowledgments}

The author would like to thank J.L. Hewett and H. Davoudiasl for discussions 
related to this work. 

\section*{References}

%
\def\MPL #1 #2 #3 {Mod.~Phys.~Lett.~{\bf#1},\ #2 (#3)}
\def\NIM #1 #2 #3 {Nucl.~Instrum.~Methods~{\bf#1},\ #2 (#3)}
\def\NPB #1 #2 #3 {Nucl.~Phys.~{\bf#1},\ #2 (#3)}
\def\PLB #1 #2 #3 {Phys.~Lett.~{\bf#1},\ #2 (#3)}
\def\PR #1 #2 #3 {Phys.~Rep.~{\bf#1},\ #2 (#3)}
\def\PRD #1 #2 #3 {Phys.~Rev.~{\bf#1},\ #2 (#3)}
\def\PRL #1 #2 #3 {Phys.~Rev.~Lett.~{\bf#1},\ #2 (#3)}
\def\RMP #1 #2 #3 {Rev.~Mod.~Phys.~{\bf#1},\ #2 (#3)}
\def\ZP #1 #2 #3 {Z.~Phys.~{\bf#1},\ #2 (#3)}
\def\IJMP #1 #2 #3 {Int.~J.~Mod.~Phys.~{\bf#1},\ #2 (#3)}

\end{document}